\documentclass[preprintnumbers,aps,prd,floatfix,nofootinbib,onecolumn]{revtex4}
\usepackage{qcd}
\usepackage{epsfig}
\usepackage{graphicx}
\usepackage{standalone}
\usepackage[normalem]{ulem}
\usepackage[pagebackref]{hyperref}
\usepackage{bm}
\usepackage{float}
\usepackage{orcidlink}
\usepackage{ifthen}
\usepackage{etoolbox}

\usepackage{tikz}
\usetikzlibrary{decorations.pathmorphing,decorations.markings,arrows.meta,calc,bending}
\definecolor{cornflowerblue}{HTML}{6495ED}

\definecolor{nicegreen}{rgb}{0.,0.5,0.}

\begin{document}

\title{What are the consequences of independent factorization and renormalization scales?}

\author{T.~C.~Rogers \orcidlink{0000-0002-0762-0275}\,}
\email{trogers@odu.edu}
\affiliation{Department of Physics, Old Dominion University, Norfolk, VA 23529, USA}

\author{R.~M.~Whitehill \orcidlink{0000-0001-5007-2473}\,}
\email{rwhit058@odu.edu}
\affiliation{Department of Physics, Old Dominion University, Norfolk, VA 23529, USA}

\begin{abstract}
 It is common for separate factorization and renormalization scales to be discussed in connection with phenomenological applications of QCD factorization theorems.  We observe that simultaneously preserving renormalization group invariance, Ward identities, and the basic sum rules in the definitions of parton densities forces these scales to be equal. The statement applies to generalized pole subtraction schemes that use dimensional regularization and to collinear factorization theorems for basic processes like deep inelastic scattering. We discuss implications for estimating the effects of scale sensitivity in phenomenological calculations, consistent extractions of Standard Model parameters alongside parton densities in global QCD analyses, and generally connecting phenomenologically extracted parton densities to first principles non-perturbative techniques like lattice QCD.
\end{abstract}

\date{\today}

\maketitle

\section{Introduction}

QCD factorization theorems involving collinear parton distribution functions (PDFs) introduce at least one auxiliary scale $\mu$ associated with the renormalization of the theory.
``Auxiliary scales'' refers to invariant mass parameters that, in principle, do not affect physical observables, but which are necessary or useful for intermediate steps of calculations. 
Consider a typical  process like deep inelastic scattering (DIS) as an example.
In ordinary minimal or modified-minimal subtraction renormalization schemes ($\text{MS}$ or \msbar, respectively) the factorization theorem allows us to express physical observables such as structure functions in the form of the convolution\footnote{For brevity, all formulas in this section are kept schematic, where for example, \eref{schematic2} is meant in the sense of an inverted convolution integral. 
We will restore flavor indices, convolutions, etc, in later sections.} \cite{Collins:2011qcdbook}
\begin{equation}
\label{e.schematic}
F(x,Q) = \mathcal{C}(x/\xi,Q/\mu;g(\mu)) \otimes f(\xi;\mu)
,\end{equation}
where $x$ and $Q$ are process-specific kinematical variables, $f(\xi;\mu)$ is a PDF, $g(\mu)$ is the renormalized strong coupling, $\mathcal{C}$ is a hard coefficient function, and $\otimes$ denotes the usual convolution integral in the longitudinal momentum fraction.
The physical observable is independent of $\mu$, 
\begin{equation}
\frac{\diff{}{F(x,Q)}}{\diff{\ln \mu}{}} = 0 \, , \label{e.schematicfact}
\end{equation}
though $\mathcal{C}(x/\xi,Q/\mu;g(\mu))$ and $f(\xi;\mu)$ individually are not.
Factorization theorems, like that in \eref{schematic}, are derived after defining PDFs in terms of matrix elements of universal operators in a fixed renormalization scheme.
In particular, in an inverted form, \eref{schematic} reads
\begin{equation}
\label{e.schematic2}
\mathcal{C}(x/\xi,Q/\mu;g(\mu)) = \frac{F(x,Q)}{f(\xi;\mu)} 
\end{equation}
so that the infrared/non-perturbative behavior on the right cancels between the observable $F(x,Q)$ and PDF $f(\xi;\mu)$, leaving the hard part sensitive only to the ultraviolet (UV) physics of the process and calculable perturbatively in QCD by exploiting asymptotic freedom.

Since QCD is a finite-distance interaction, the only divergences present in fully non-perturbative treatments of PDFs are the ordinary UV divergences associated with renormalization. 
Therefore, the renormalization scale $\mu$ is the only auxiliary scale necessary for defining a PDF. 
Because the left side of \eref{schematic2} inherits all its auxiliary scale dependence from that of the PDF, $\mu$ is also the only auxiliary scale that appears in the hard part $\mathcal{C}$.
Once a UV renormalization scheme is fixed for the PDF, the hard part is completely determined.

Hence, the structure of factorization leads most naturally to a single necessary auxiliary scale. 
It has sometimes been found useful, however, to introduce a separate, additional auxiliary scale, called a factorization scale (denoted by $\mu_F$ generally), for calculating individual Feynman diagrams in massless perturbative QCD.
This provides a heuristic parameter for demarcating hard and collinear momentum regions in Feynman diagrams for on-shell and massless partons\footnote{See~\cite[Sec.~3.2]{CTEQ:1993hwr},~\cite[Sec.~2]{Martin:2017uyr}, and~\cite{yadism-scale-variations} for some examples of how the factorization scale terminology is typically used. We discuss this more in \sref{factscale}}. 
The factorization scale is often treated as conceptually separate from the UV renormalization scale $\mu$, and thus is taken to be independently adjustable.
In this paper, we study the impact that doing this has on the general properties of parton densities at the level of their basic operator definitions.

Fundamental properties of PDFs, such as the quark number and momentum sum rules, follow from their operator definitions and play important roles as constraints in phenomenological applications.
It is therefore critical to verify that they are not spoiled by the details of renormalization. While most of the basic sum rules are known to hold in the commonly used $\text{MS}$ and $\msbar$ schemes, the possible impact of introducing extra auxiliary parameters like $\mu_F$ has not, to our knowledge, been addressed in detail in the literature within this context.

These issues grow more pressing as new possibilities arise for performing first-principles calculations of PDFs directly starting from non-perturbative QCD, and for comparing those calculations with phenomenological extractions \cite{Lin:2017snn,Cichy:2018mum,Constantinou:2020hdm}.
In that context, the details of the operator definition, including its interpretation and its general properties, become central.
Constraints like number and momentum sum rules strengthen the predictive power of QCD factorization calculations, whereas the presence of many independently adjustable auxiliary scales, each acting as a separate potential source of uncertainty, may weaken predictive power.
For example, it was pointed out recently in Ref.~\cite{Delorme:2025teo} that nonstandard renormalization schemes can modify sum rules through higher order corrections, and prescriptions are needed to take this into account when switching between schemes~\cite{Delorme:2026vln}. 
The effect of scheme choices can also impact the non-perturbative modeling of input PDFs in global analyses, as discussed in Ref.~\cite{Courtoy:2020fex}. Understanding the degree to which theoretical properties of PDFs are established, and the sense in which they can act as constraints on parton correlation functions~\cite{Whitehill:2026mrr}, is especially important in light of ``big data paradoxes''~\cite{meng2018statistical,Courtoy:2022ocu} facing QCD phenomenology.
Here, we argue that simultaneously preserving exact PDF sum rules, gauge invariance, and RG invariance generally requires the use of only one independent auxiliary scale $\mu$. 
In other words, introducing multiple adjustable scales complicates the preservation of symmetries at the level of PDFs in light of RG invariance at the level of the full QCD Lagrangian.

The paper is organized as follows:
In \sref{cons}, we review the basics of renormalization and its connection to conservation laws.
We then review the renormalization of PDFs in \sref{pdfs} and contrast it with treatments that incorporate the concept of a factorization scale in \sref{factscale}.
In \sref{sumrules}, we relate sum rule derivations for a renormalized PDF with past derivations that start with a factorization scale included and explain how the latter generally conflicts with RG invariance.
We then conclude with some discussion in \sref{pheno} about the phenomenological implications of our findings and a brief summary and outlook in \sref{conclusions}.

\section{Conservation laws and renormalization}
\label{s.cons}

PDF sum rules encode universal conservation laws of the underlying quantum field theory (\text{e.g.} QCD), independent of any specific process. 
To preserve those conservation laws, and therefore the sum rules, under renormalization, the cancellation of auxiliary scale dependence will only involve those scales associated with the renormalization of the theory itself. 
Thus, if a PDF is defined to depend non-trivially on an extra auxiliary scale, then sum rules associated with it will generally require corrections, reflecting the fact that they will no longer exactly encode the basic conservation laws from which they are derived at the level of the PDF definition.
To correct this, the extra scale dependence might be incorporated into the renormalization of the theory itself at the level of the Lagrangian as a second renormalization scale.
However, with the extra scale promoted to an RG scale, RG invariance, along with any other symmetry principles, constrain its relationship with other RG scales and the parameters of the theory.
In this paper, we make these observations concrete through a series of explicit examples.

Because our results are meant to apply generally to any finite-range renormalizable quantum field theory, we organize the discussion around a massive Yukawa theory.
Calculations can then be performed explicitly across all relevant scales using standard perturbation theory, even in the infrared regions.
The theory is defined in $D=4-2\epsilon$ dimensions, using dimensional regularization to regulate UV divergences and keeping $\epsilon > 0$ until the end of all calculations, by the bare Lagrangian density
\begin{align}
\mathcal{L} &{}= i \overline{\psi}_0 \sla{\partial} \psi_0 - m_{q,0} \overline{\psi}_0 \psi_0 + \frac{1}{2} \partial_\mu \phi_0 \partial^\mu \phi_0 - \frac{m_{s,0}^2}{2} \phi_0^2 - g_0 \, \overline{\psi}_0 \psi_0 \phi_0 + V(\phi_0) \, . 
\label{e.thelagrange}
\end{align}
We abbreviate the three- and four-point scalar self-interactions that are needed to make the theory fully renormalizable by $V(\phi_0)$, although they do not contribute at the perturbative order considered in this work. 
We introduce renormalized fields through
\begin{equation}
\begin{aligned}
\phi &\equiv Z_s^{-1/2} \phi_0 \, , \\
\psi &\equiv Z_q^{-1/2} \psi_0 \, .
\end{aligned}
\label{e.renorms}
\end{equation}
so that the Lagrangian becomes
\begin{align}
\mathcal{L} &{}= i Z_q \overline{\psi} \sla{\partial} \psi - Z_{q} m_{q,0} \overline{\psi} \psi + \frac{Z_s}{2} \partial_\mu \phi \partial^\mu \phi - \frac{Z_{s} m_{s,0}^2}{2} \phi^2 - Z_{q} Z_{s}^{1/2} g_0 \, \overline{\psi} \psi \phi + V(Z_s^{1/2} \phi) \, ,
\label{e.thelagrangerenorm}
\end{align}
We then define counterterms and the renormalized masses and couplings through
\begin{equation}
\begin{gathered}
\begin{aligned}
Z_s &\equiv 1 + \delta Z_s \, ,
&
Z_s m_{s,0}^2 &\equiv m_s^2 + \delta m_s^2 \, , \\
Z_q &\equiv 1 + \delta Z_q \, ,
&
Z_q m_{q,0} &\equiv m_q + \delta m_q \, ,
\end{aligned}
\\
Z_q Z_s^{1/2} g_0 \equiv \mu^{\epsilon} g + \delta g \, .
\end{gathered}
\label{e.Zs}
\end{equation}
This allows \eref{thelagrangerenorm} to be separated into a renormalized and counterterm Lagrangian respectively,
\begin{align}
\label{e.renormlag}
\begin{aligned}
\mathcal{L} &{}= i \overline{\psi} \sla{\partial} \psi - m_{q} \overline{\psi} \psi + \frac{1}{2} \partial_\mu \phi \partial^\mu \phi - \frac{m_{s}^2}{2} \phi^2 - \mu^\epsilon g  \overline{\psi} \psi \phi + V(\phi) \, \\
&{}+ i \delta Z_q \overline{\psi} \sla{\partial} \psi + \frac{\delta Z_s}{2} \partial_\mu \phi \partial^\mu \phi - \delta m_{q} \overline{\psi} \psi - \frac{\delta m_{s}^2}{2} \phi^2 - \delta g  \overline{\psi} \psi \phi + \delta V(\phi) \, .
\end{aligned}
\end{align}
The prescription for assigning values to $\delta Z_q$, $\delta Z_s$, $\delta m_q$, $\delta m_s^2$, $\delta g$, and $\mu$ defines the renormalization scheme.
In a generalized minimal subtraction scheme, the counterterms are defined to subtract out only those terms proportional to powers of $A^\epsilon/\epsilon$ where $A$ is a dimensionless numerical factor,
\begin{subequations}
\begin{align}
 \delta Z_s(g(\mu),\epsilon) &= \sum_{n=1}^\infty \parz{\frac{A^\epsilon}{\epsilon}}^n C^{Z_s}_n(g(\mu)) \label{e.ct1} \\
 \delta Z_q(g(\mu),\epsilon) &= \sum_{n=1}^\infty \parz{\frac{A^\epsilon}{\epsilon}}^n C^{Z_q}_n(g(\mu))  \label{e.ct2} \\
 \delta m_s^2(g(\mu),\epsilon) &= m_s^2  \sum_{n=1}^\infty \parz{\frac{A^\epsilon}{\epsilon}}^n C^{m_s^2}_n(g(\mu)) \label{e.ct3} \\
 \delta m_q(g(\mu),\epsilon) &= m_q  \sum_{n=1}^\infty \parz{\frac{A^\epsilon}{\epsilon}}^n C^{m_q}_n(g(\mu)) \label{e.ct4} \\
 \delta g(g(\mu),\epsilon) &=  \mu^\epsilon g \sum_{n=1}^\infty \parz{\frac{A^\epsilon}{\epsilon}}^n C^g_n(g(\mu)) \label{e.ct5}
\end{align}
\label{e.cts}
\end{subequations}
The $C_n$'s are $\epsilon$-independent coefficients determined from perturbation theory. 
Ordinary \text{MS} corresponds to $A = 1$ while $\msbar$ corresponds to $A = 4 \pi e^{-\gamma_E}$. 
Keeping an $A \neq 1$ amounts to including a $\ln A$ with each $1/\epsilon$ pole that is subtracted in a calculation.
The renormalization scheme only becomes fully specified after we assign numerical values to both $A$ and $\mu$.

The bare theory in \eref{thelagrange} depends on the UV regulator $\epsilon$ but is independent of quantities like $\mu$, which are associated only with the renormalization scheme choice that separates the Lagrangian into renormalized and counterterm parts in \eref{renormlag}.
This gives the usual RG invariance relations like 
\begin{equation}
\label{e.g0running}
\frac{\diff{}{g_0}}{\diff{\ln \mu}} = 0 \, .
\end{equation}

Since it is the bare operators that satisfy the canonical commutation relations, many basic properties are most naturally established in the bare theory.
By contrast, most calculations relevant to physical measurements use renormalized quantities in four dimensions, so it is useful to identify those properties which carry over directly from the bare to the renormalized theory.
Consider, for example, the expectation value of the conserved current $j^\mu_0(0) = \overline{\psi}_0(0) \gamma^\mu \psi_0(0)$, corresponding to the flow of charge,
\begin{equation}
\langle P |  j^\mu_0(0) | P \rangle = \langle P | \overline{\psi}_0(0) \gamma^\mu \psi_0(0) |P \rangle = 2 P^\mu Q_P \, ,  \label{e.chargecons}
\end{equation}
where $|P \rangle$ denotes a momentum and charge eigenstate carrying four-momentum $P^{\mu}$ and
charge $Q_P$. 
In QCD, where there are multiple quark flavors, each quark field has a corresponding conserved current of the form \eref{chargecons}, and our discussion below applies independently to each such current.

Conservation laws like \eref{chargecons} continue to hold exactly after renormalization.
Indeed, there is no renormalization of such currents, and the cancellation of  UV divergent behavior in the relevant Feynman diagrams is accounted for by counterterms in the Lagrangian alone. Here and below, we focus on \text{MS} and $\msbar$, since these are the schemes used in our later examples\footnote{The results are not exclusive to dimensional regularization with minimal subtraction schemes, however. 
Analogous conclusions hold in other renormalization schemes like the BPHZ scheme that preserve the relevant symmetries and satisfy the assumptions underlying the corresponding non-renormalization theorems}.
Such schemes remove only the divergent parts of unsubtracted quantities that become ill-defined when the UV regulator is removed, and all finite renormalized quantities therefore inherit exactly the symmetry and conservation properties of the bare theory.

However, as discussed in more detail in \cite[Sec. 6.6, pg. 151]{Collins:1984xc}, these observations, including non-renormalization theorems for relationships like \eref{chargecons}, rely on assumptions about the renormalization scheme. More general renormalization schemes may violate these assumptions so that such conclusions do not hold. 
In the context of PDFs, 
sum rules and other operator constraints are consequences of underlying conservation laws, such as the current conservation relation \eref{chargecons}.
Others involve equations of motion and further identities and related symmetries.
Introducing auxiliary scales such as $\mu_F$, in addition to the renormalization scale $\mu$, does more than simply rescale the common subtraction constant $A$ that relates ordinary \text{MS}-type conventions.
It constitutes a non-trivial multiscale generalization of the usual \text{MS} or $\msbar$ schemes, and hence is a formally distinct renormalization scheme.

\section{Parton density renormalization}
\label{s.pdfs}

In this section, we briefly review the ordinary renormalization of $\msbar$ quark parton densities. 
Details are found in section 8.7 of Ref.~\cite{Collins:2011qcdbook}.
The operator definition for a bare unpolarized quark-in-quark PDF is
\begin{equation}
f_{q/q,0}(\xi;\epsilon) = \int \frac{\diff{y^-}}{2 \pi} e^{-i \xi P^+ y^-} \langle P | \overline{\psi}_0(0,y^-,\T{0}{}) \frac{\gamma^+}{2} \psi_0(0) | P \rangle \, , \label{e.barepdf}
\end{equation}
where the light-cone momentum fraction $\xi = k^+ / P^{+}$ denotes the ratio of the struck and parent quark plus components of momenta, $k^{+}$ and $P^{+}$, respectively.
In \eref{barepdf}, the state $|P\rangle$ denotes an asymptotic on-shell quark state from the Yukawa theory in \eref{thelagrange}.
Because we keep non-zero masses in \eref{renormlag}, the bare PDF has no infrared or collinear divergences, but it does have UV divergences in four dimensions. 
The statement that the PDF is renormalizable means that a finite, four-dimensional distribution can be defined as a convolution of the bare PDF $f_{0}(\xi;\epsilon)$ with a renormalization factor $Z^\text{pdf}(g(\mu),\xi,\epsilon)$, independent of the large-distance structure of the external state $|P \rangle$, including the mass scales.
The renormalized PDF is then given by\footnote{Note that the integral can run to $\xi' < \xi$ because the bare PDF vanishes for $\xi/\xi' > 1$.}
\begin{equation}
 f_{q/q}(\xi;\mu;\epsilon) =  Z_{qj}^\text{pdf}(g(\mu),\xi',\epsilon) \otimes f_{j/q,0}(\xi/\xi';\epsilon) = \int_0^1 \frac{\diff{\xi'}{}}{\xi'} Z_{qj}^\text{pdf}(g(\mu),\xi',\epsilon) f_{j/q,0}(\xi/\xi';\epsilon) \, , \label{e.renpdf}
\end{equation}
where repeated flavor indices $j$ are summed over.
PDFs used in phenomenological applications are obtained by setting $\epsilon \to 0$.
The RG equation for $f(\xi;\mu;\epsilon)$ (the DGLAP equation~\cite{Altarelli:1977zs,Dokshitzer:1977sg,Gribov:1972ri}) follows from \eref{renpdf} and the RG invariance of the bare PDF,
\begin{equation}
\frac{\diff{}}{\diff{\ln \mu}{}} f_{j/q,0}(\xi;\epsilon) = 0 \, . \label{e.pdfrg}
\end{equation}
The running of the renormalized PDF that follows from \eref{pdfrg} is analogous to the running of the renormalized coupling that follows from \eref{g0running}.
For doing explicit calculations, it is often useful to rewrite \eref{barepdf} in terms of renormalized fields,  
\begin{equation}
f_{q/q,0}(\xi;\epsilon) = Z_q \int \frac{\diff{y^-}}{2 \pi} e^{-i k^+ y^-} \langle P | \overline{\psi}(0,y^-,\T{0}{}) \frac{\gamma^+}{2} \psi(0) | P \rangle \equiv Z_q f_{q/q}^\text{unsub}(\xi;\mu;\epsilon)  \, , \label{e.barepdfunsub}
\end{equation}
and to reexpress the renormalization in \eref{renpdf} as
\begin{equation}
 f_{q/q}(\xi;\mu;\epsilon) =  Z_q Z_{qj}^\text{pdf}(g(\mu),\xi',\epsilon) \otimes f^{\text{unsub}}_{j/q}(\xi/\xi';\mu;\epsilon) \, . \label{e.frenorm}
\end{equation}
Since it is defined with renormalized fields, $f^{\text{unsub}}_{j/q}(\xi;\mu;\epsilon)$ is the correlation function calculated before $Z_q Z_{qj}^\text{pdf}$ counterterms are subtracted, represented graphically by diagram (a) in \fref{qq_sq_defs}.
In a minimal subtraction renormalization scheme, the counterterm expansion is
\begin{align}
\left[ Z_q Z_{qj}^\text{pdf} \right](g(\mu),\xi,\epsilon) &= \delta_{qj} \delta(1-\xi) + \delta \left[ Z_q Z_{qj}^\text{pdf} \right] (g(\mu),\xi,\epsilon) = \delta_{qj} \delta(1-\xi) +  \sum_{n=1}^\infty \parz{\frac{A^\epsilon}{\epsilon}}^n C^{Z^{\text{pdf}}}_{qj,n}(g(\mu),\xi) \label{e.ct6}
,\end{align}
to match the other counterterms in \eref{cts}.
Note that we have written the counterterm expansion in \eref{ct6} with an explicit $Z_q$ included since it is the combination of $Z_q$ and $Z_{qj}^\text{pdf}$ that appears in \eref{frenorm}.

\begin{figure}[t]
    \centering
\includegraphics[width=7.0cm]{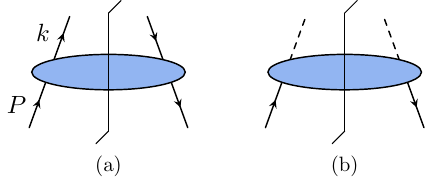}
    \caption{Generic Feynman diagrams corresponding to (a) $f_{q/q}^{\rm unsub}$ and (b) $f_{s/q}^{\rm unsub}$.}
    \label{f.qq_sq_defs}
\end{figure}

The sum over $j$ in equations like \eqref{e.frenorm} includes all types of partons, so we will also need the bare antiquark-in-quark PDF $f_{\bar{q}/q,0}(\xi;\epsilon)$, obtained by taking $\overline{\psi} \leftrightarrow \bar{\psi}$ in \eref{barepdf}, and the scalar-in-quark PDF, 
\begin{equation}
f_{s/q,0}(\xi;\epsilon) = Z_s \xi P^+ \int \frac{\diff{y^-}}{2 \pi} e^{-i k^+ y^-} \langle P | \phi(0,y^-,\T{0}{}) \phi(0) | P \rangle \equiv Z_s f_{s/q}^\text{unsub}(\xi;\mu;\epsilon)  \, , \label{e.barepdfunsubsca}
\end{equation}
where the unsubtracted scalar-in-quark PDF is graphically represented by diagram (b) in \fref{qq_sq_defs}.
The renormalization of $f_{s/q,0}$ is analogous to \eref{renpdf} with renormalization factors $Z_{sj}^\text{pdf}$, 
\begin{equation}
 f_{s/q}(\xi;\mu;\epsilon) =  Z_{sj}^\text{pdf}(g(\mu),\xi',\epsilon) \otimes f_{j/q,0}(\xi/\xi';\mu;\epsilon) \, , \label{e.renPDFsca}
\end{equation}
and with an expansion of counterterms directly analogous to \eref{ct6}. 

In the most general version of minimal subtraction, we could allow the $A$ constant in each counterterm to be different. 
Thus, for example, we might write 
\begin{equation}
A \to A_{m_s} \, , \;\; A \to A_{m_q} \, , \; \; A \to A_{q} \, , \; \; A \to A_{s} \, , \; \; A \to A_{\text{pdf},qj} \label{e.genMSreplace}
\end{equation}
in \erefs{ct2}{ct5} and \eref{ct6} respectively. 
We follow Ref.~\cite[pg.~285]{Sterman:1993hfp} and refer to this as a ``generalized minimal subtraction'' scheme. 
These generalized $A$ constants can be written as ratios of a mass scale with $\mu$.
For example, $A_{\text{pdf},qj} = (\mu^2/\mu_F^2) \!\times\! A$ so that the effect is equivalent to introducing multiple invariant mass scales into the scheme beyond just the $\mu$ in \eref{renormlag}.
Ultimately, doing this simply makes some logarithms of $\mu$ into logarithms of $\mu_F$ and any additional scales from the other $A$ constants.

The counterterms in \eref{cts} and \eref{ct6} cannot be chosen completely independently of one another if properties of the bare theory like sum rules are to be preserved under renormalization.
Because bare quantities are RG invariant, the divergent UV pole structure of $f_{q/q}^{\text{unsub}}$ in \eref{frenorm}, which is calculated with renormalized fields and parameters like $g(\mu)$, is related to the field, mass, and coupling counterterms in the Lagrangian.
The full PDF counterterm $Z_q Z_{qj}^\text{pdf}$ in \eref{ct6}, which is constructed from the same renormalized parameters, must cancel this UV structure.
In particular, these cancellations are responsible for the non-renormalization of currents such as \eref{chargecons}.
Thus, there is a consistency constraint between $Z_q Z_{qj}^\text{pdf}$ and the other renormalization counterterms. 
Changes in the scheme for regular RG counterterms in the Lagrangian and the scheme for the PDF counterterms cannot be made independently of one another if RG invariance and fundamental symmetries are to be preserved simultaneously at the PDF level.
We give further illustrations involving PDFs in \sref{sumrules} and \sref{sumrulefail}.

\section{factorization scales}
\label{s.factscale}

Our purpose with this paper is to extend the discussion above to incorporate the concept of a factorization scale, and to compare standard minimal subtraction schemes with multi-scale minimal subtraction schemes like in \eref{genMSreplace}. In this section, therefore, we elaborate further on the question of what a  ``factorization scale'' specifically refers to. 
We caution that the term has many meanings in the literature, but for our purposes it is meant only in the basic sense that is common in textbook treatments like Refs.~\cite{Ellis:1996mzs,Dissertori:2003pj,Devenish:2004pb,Campbell:2017hsr}.
In that context, the introduction of a factorization scale is motivated by the observation that, within specific processes like DIS, extending the naive parton model to include perturbative higher order parton emissions results in collinear divergences from massless, unregulated Feynman diagram calculations.
For example, a typical low order calculation of a partonic DIS structure function $\widehat{F}(\xi,Q)$ for an on-shell massless target takes the schematic form
\begin{equation}
\widehat{F}(\xi,Q) \sim A^\epsilon g^2 \mu^{2 \epsilon} \int_0^{Q^2} \frac{\diff{\Tscsq{k}{}}{}}{\Tscsq{k}{}} (\Tscsq{k}{})^\epsilon [\hdots]  \sim g^2 \parz{\frac{1}{\epsilon} + \ln A} +g^2 \ln \frac{Q^2}{\mu^2} + \hdots \, , \label{e.factscale1}
\end{equation}
where 
$\Tsc{k}{}$ is the transverse momentum of a parton emission. 
This style of setup, and its contrast with ones centered on factorization derivations with renormalized PDFs, is discussed in much more detail in Sec. III of Ref.~\cite{Collins:2021vke}.
The external, process-specific scale $Q^2$ imposes a natural UV cutoff on large transverse momentum, so that rather than finding UV divergences as we did in the treatment of PDF operator matrix elements in \sref{pdfs}, the $1/\epsilon + \ln A$ here 
appears because of collinear divergences that result from integrating $\Tscsq{k}{}$ into the non-perturbative region around $\Tsc{k}{} \approx 0$. 
Confronting the problem of factorization from this perspective, the divergence appears to be associated with a regulator on small partonic $\Tscsq{k}{}$ rather than a scale associated with UV renormalization.
Without the clear connection to UV renormalization, there seems to be an extra freedom to introduce a factorization scale $\mu_F \ne \mu$ for the purpose of marking off collinear partons.
That freedom gets incorporated into perturbative calculations by rewriting expressions like \eref{factscale1} as 
\begin{align}
\widehat{F}(\xi,Q) \sim g^2 \parz{\frac{1}{\epsilon} + \ln A + \ln \frac{\mu_F^2}{\mu^2}}  + g^2 \ln \frac{Q^2}{\mu_F^2} + \hdots \, . \label{e.factscale2}
\end{align}
Finally, the first term is absorbed into a redefinition of the PDF. 
It is through methods of derivation like these that extra auxiliary scales like $\mu_F$ enter into formulations of factorization.

In some versions of this style of setup, an explicit lower cutoff $\kappa$ is used to regulate the collinear divergence instead of dimensional regularization. 
Doing that emphasizes an interpretation of the divergence as arising from collinear as opposed to UV partons.
Then, \eref{factscale1} is written as 
\begin{equation}
\widehat{F}(\xi,Q) \sim  g^2 \int_{\kappa^2}^{Q^2} \frac{\diff{\Tscsq{k}{}}{}}{\Tscsq{k}{}} [\hdots]  \sim g^2 \ln \frac{Q^2}{\kappa^2} + \hdots \, , \label{e.factscale1p}
\end{equation}
and \eref{factscale2} becomes 
\begin{align}
\widehat{F}(\xi,Q) \sim  g^2 \parz{\ln A + \ln \frac{\mu_F^2}{A \kappa^2}} + g^2 \ln \frac{Q^2}{\mu_F^2} + \hdots \, , \label{e.factscale2p}
\end{align}
where the collinear divergence now comes from setting the collinear regulator $A \kappa^2$ to zero rather than from setting $\epsilon \to 0$ in the $1/\epsilon$ pole as in \eref{factscale2}.

In approaches that follow treatments like the above, the next step in obtaining a renormalized PDF is to absorb the collinear divergent contributions in \eref{factscale1} and \eqref{e.factscale2p} into updated PDF definitions.
Since the PDFs in this approach are typically treated as purely phenomenological inputs rather than objects computed from first principles, this does not create any obvious problems, and iterating the procedure order-by-order absorbs all the collinear $1/\epsilon$ poles. 
When $\mu_F \ne \mu$, the accompanying $\ln(\mu^2/\mu_F^2)$ terms can also be absorbed into the PDFs.
In practice, this process of absorbing collinear poles is equivalent to adding the generalized-\text{MS} UV counterterms of \eref{genMSreplace} in the renormalized operator-defined PDFs of \sref{pdfs}.
Translated into the treatment of renormalization from \sref{pdfs}, the effect on the renormalization of PDFs is to modify the basic $\msbar$-renormalized definition.
More concretely, the renormalization counterterm for the unsubtracted quark-in-quark PDF in \eref{ct6} is modified to read
\begin{align}
\delta \left[ Z_q Z_{qj}^\text{pdf} \right](g(\mu),\xi,\epsilon,\mu_{\text{pdf},qj}/\mu) = 
\sum_{n=1}^\infty \parz{\frac{A_{\text{pdf},qj}^\epsilon}{\epsilon}}^n C^{Z^{\text{pdf}}}_{qj,n}(g(\mu),\xi) \, , \label{e.ct6p}
\end{align}
with
\begin{equation}
A_{\text{pdf},qj} = A \frac{\mu^2}{\mu_{\text{pdf},qj}^2} \, , \label{e.Arep}
\end{equation}
and similarly for the scalar-in-quark PDF.
Hence, the structure of the counterterm is the same as in ordinary $\text{MS}$ or $\msbar$ schemes but with a modified ``$A$'' for the PDF renormalization factor, where $\mu_{\text{pdf},qj}$ in \eref{Arep} plays the same role as the $\mu_F$ from \eref{factscale2}.
In principle, we can generalize minimal subtraction still further to \eref{genMSreplace}.
For the wavefunction renormalization, we would then write
\begin{subequations}
\label{e.ctp}
\begin{align}
\delta Z_s(\mu,\epsilon;\mu_s/\mu) &=
\sum_{n=1}^\infty \parz{\frac{A_s^\epsilon}{\epsilon}}^n C^{Z_s}_n(g(\mu)) \, , \label{e.ct4p} \\
\delta Z_q(\mu,\epsilon;\mu_q/\mu) &= 
\sum_{n=1}^\infty \parz{\frac{A_q^\epsilon}{\epsilon}}^n C^{Z_q}_n(g(\mu))  \, , \label{e.ct5p}
\end{align}
\end{subequations}
with $A_q = (\mu^2/\mu_q^2) \!\times\! A$ and $A_s = (\mu^2/\mu_s^2) \!\times\! A$. 
The effect is to shift scales that appear in the wavefunction renormalization from $\mu$ to different scales $\mu_q$ and $\mu_s$.
When parton masses are relevant, similar modifications to \erefs{ct2}{ct3} can be made as well. When all replacements in \eref{genMSreplace} are made, only the coupling in \eref{ct5} involves the original $A$.

The view of factorization embodied by equations like \eref{factscale1} captures important physical intuition. 
It shows how perturbative QCD radiation logarithmically violates Bjorken scaling as the DIS phase space grows with $Q$. 
Furthermore, the steps outlined above provide a simple route to many standard perturbative results.
However, the derivation elevates the significance of unphysical infrared-divergent behavior of perturbative calculations in the collinear region where $\Tsc{k}{} \approx 0$.
Taken too literally, this can allow artifacts of the collinear regulator to propagate into final results, including into perturbatively calculable hard factors that should be insensitive to the treatment of soft, infrared physics.
Moreover, recasting UV counterterms as collinear subtractions can hide the RG constraints that relate those counterterms to one another, and this can have practical consequences when enumerating the general properties of renormalized PDFs.
Our examples in \sref{sumrulefail} illustrate this.

Another way that a factorization scale may be incorporated is through the definition of the collinear PDF by imposing $\mu_F$ as a direct UV cutoff on the $\Tsc{k}{}$-integral of a renormalized transverse momentum dependent PDF, instead of through renormalization with \erefs{barepdf}{renpdf}.
In this view, the factorization scale is then associated with UV divergences, but not through renormalization, and while calculations utilizing this definition produce the same extra $\sim g^2 \ln (\mu^2/\mu_F^2)$ terms, analogous to what appear in \eref{factscale2}, it also introduces additional terms and complications, including nonlogarithmic leading power corrections, inhomogeneous terms in the evolution, subleading corrections, and problems associated with the cancellation of lightcone divergences~\cite{Collins:2003fm}.

The existing literature contains many minor variations of the style of argument in \erefs{factscale1}{factscale2}, and there is a wide variety of different notational conventions.
In some cases, the renormalization scale is given its own subscript, as with the $\mu_r$ in Ref.~\cite{CTEQ:1993hwr}. 
In other cases, such as in Ref.~\cite[Eq.~(4.78)]{Ellis:1996mzs}, it is the factorization scale that is represented by an unsubscripted $\mu$.
Ref.~\cite{Candido:2023ujx} uses $r$ and $f$ subscripts to label renormalization and factorization scales, respectively.  
Regardless of these conventions, what is relevant for our purposes is that all approaches examined introduce an extra factorization scale and generate the same $(\mu^2/\mu_F^2)$-type logarithms that appear in \eref{ct6p}. 

\section{Sum rules}
\label{s.sumrules}

In this section we review the quark number sum rule, focusing on the implications of inserting extra factorization scales in the manner described in the previous section. 
We base our discussion on the treatment of PDFs in \cite[Sec.~4.3]{Collins:1981uw} and \cite[Chapt.~6]{Collins:2011qcdbook}. 

\subsection{Derivations}
\label{s.sumrulesa}

We start with the quark number sum rule for the bare PDF in \eref{barepdf}.
Because the relevant operator identity and canonical anticommutation relations are formulated most directly in terms of bare fields, the sum rule follows first at the bare level:
\begin{equation}
\int_0^1 \diff{\xi}{} \left[ f_{q/q,0}(\xi;\epsilon) - f_{\bar{q}/q,0}(\xi;\epsilon) \right] = 1 \, . \label{e.sumrule}
\end{equation}
Its derivation follows by integrating \eref{barepdf} and the corresponding expression for $f_{\bar{q}/q,0}(\xi;\epsilon)$ over all $\xi$. 
Using the support, $-1 \leq \xi \leq +1$, for the bare quark PDFs, and the relation $f_{q/q}(-\xi;\epsilon) = -f_{\bar{q}/q}(\xi;\epsilon)$ gives
\begin{equation}
\begin{aligned}
 \int_0^1 \diff{\xi}{} \left[ f_{q/q,0}(\xi;\epsilon) - f_{\bar{q}/q,0}(\xi;\epsilon) \right] &= 
 \int_{-\infty}^\infty \diff{\xi}{} \int \frac{\diff{y^-}}{2 \pi} e^{-i \xi P^+ y^-} \langle P | \overline{\psi}_0(0,y^-,\T{0}{}) \frac{\gamma^+}{2} \psi_0(0) | P \rangle \\
 & = \frac{1}{2 P^+}  \langle P | \overline{\psi}_0(0) \gamma^+ \psi_0(0) | P \rangle \, . 
\end{aligned}
 \label{e.baresumrule}
\end{equation}
In the last line we utilized the $2 \pi \delta(y^- P^+)$ that comes  from the $\xi$ integration.
Taking the quark number $Q_P=1$ for the state $| P \rangle$, we recover \eref{sumrule} from \eref{chargecons}.
This style of derivation makes it clear how the sum rule for the bare PDF is essentially a restatement of the local conservation of quark number in the bare theory. 

Next we obtain the same sum rule for the renormalized PDF in ordinary minimal subtraction, using a single universal subtraction constant $A$ in \eref{cts} and \eref{ct6}.
To do this, we show that the non-renormalization theorem for the current in \eref{chargecons}, applied directly to the local operator, remains valid when the current is obtained by integrating the renormalized PDF. 
That is, we show that the integration in \eref{baresumrule} commutes with renormalization. 
Applying the $\xi$ integral from $-\infty$ to $\infty$ to \eref{renpdf} gives
\begin{equation}
\begin{aligned}
 \int_{-\infty}^\infty \diff{\xi}{} f_{q/q}(\xi;\mu;\epsilon) & = \int_0^1 \frac{\diff{\xi'}{}}{\xi'} Z_{qj}^\text{pdf}(g(\mu),\xi',\epsilon) \int_{-\infty}^\infty \diff{\xi}{}  f_{j/q,0}(\xi/\xi';\epsilon) 
 = 
 \parz{\int_0^1 \diff{\xi'}{} Z_{qj}^\text{pdf}(g(\mu),\xi',\epsilon)} \delta_{jq} \, 
 ,\label{e.renpdf3}
\end{aligned}
\end{equation}
where the second equality follows from the substitution $\xi = \xi''/\xi'$ and the sum rule for the bare PDFs.
Now we verify that the remaining integral in \eref{renpdf3} reduces to
\begin{align}
\int_0^1 \diff{\xi'}{} Z_{qj}^\text{pdf}(g(\mu),\xi',\epsilon) = \delta_{qj} \, \label{e.Zint}
\end{align}
in ordinary minimal subtraction.
To see this, note the following: {\bf i)} the left side of \eref{renpdf3} is finite and independent of the UV regulator $\epsilon$ when $\epsilon \to 0$; {\bf ii)} for $A=1$, the only term on the right side of \eref{ct6} that is finite and $\epsilon$-independent as $\epsilon \to 0$ is the first term in \eref{ct6} corresponding to $\delta_{qj} \delta(1-\xi)$, while $\delta [Z_q Z_{qj}^{\rm pdf}]$ must integrate to zero since it contains purely the $1/\epsilon$ poles multiplied by $\epsilon$-independent coefficients; and {\bf iii)} for $A \ne 1$, we may simply rescale the value of $\mu$ from the $A=1$ case, which does not impact the logic of steps (i) and (ii).
These observations guarantee that \eref{Zint} holds when a single constant $A$ is used for all renormalization counterterms in \eref{cts} and \eref{ct6}, and
therefore, \eref{renpdf3} carries over directly under renormalization,
\begin{align}
& \int_0^1 \diff{\xi}{} \left[ f_{q/q}(\xi;\mu;\epsilon) - f_{\bar{q}/q}(\xi;\mu;\epsilon) \right] = \int_{-\infty}^{\infty} \diff{\xi}{} f_{q/q}(\xi;\mu;\epsilon) 
 = 1 \, .
 \label{e.renpdf4}
\end{align}
so that the quark number sum rule is preserved in renormalization schemes such as $\text{MS}$ and $\msbar$.
We see that it is essentially a version of the non-renormalization theorem for the quark number current in \eref{chargecons}.
Similar arguments can be made for other important sum rules such as the momentum sum rule, 
\begin{equation}
\label{e.momrule}
\sum_j \int_0^1 \diff{\xi} \xi f_{j/q}(\xi;\mu) = 1 \, .
\end{equation}
where the sum is over all parton species and flavors.
In the case of the momentum sum rule, the Noether current is a component of the stress-energy tensor~\cite[Sec.~4.3]{Collins:1981uw}. 

Arguments like that above for the validity of \eref{Zint} do not necessarily hold in more general renormalization schemes. 
In various extensions of generalized $\text{MS}$ like those described in \sref{factscale}, complications can arise 
from the introduction of separate factorization scales where there are different distinct values for $A$ in different counterterms.
To see why, recall that the form of $Z_{qj}^\text{pdf}$ in \text{MS} is determined by 
both the value of $A$ in \eref{ct6} and the form of the other renormalization factors in \eref{cts}.
It is the renormalized parameters and Lagrangian counterterms that are used to calculate $f_{q/q}^{\rm unsub}$ and 
which ensure that its divergent part is simply a series of $A^\epsilon/\epsilon$ poles with a common $A$, corresponding to a single value of $\mu$.
In the case that different $A$'s are used for the different counterterms, it should be expected that higher-order, finite and non-zero terms modify the right side of \eref{Zint}.

\subsection{Other derivations}

In past literature it is possible to find derivations of the sum rules in \erefs{renpdf4}{momrule} within multi-scale generalized-\text{MS} schemes~\cite{Beenakker:2015rna}. 
These derivations follow a ``track-B'' approach to factorization, as described in~\cite{Collins:2021vke}, and to which we alluded in \sref{factscale}.
In that approach, the starting point is to postulate a factorization formula (for example, in DIS) of the form\footnote{For this subsection, we revert back to the schematic notation like that in \eref{schematic}, \textit{e.g.} with flavor indices and arguments left implicit.}
\begin{equation}
\label{e.trackB}
F(x,Q^2) = \widehat{F}^{\text{partonic}} \otimes f^{\text{bare},B} \, , 
\end{equation}
where $\widehat{F}^{\text{partonic}}$ is an unsubtracted structure function for scattering off a partonic target.\footnote{By ``unsubtracted'' we mean here that it is a full partonic structure function calculated in massless perturbation theory including the collinearly divergent poles.}
(We have included a ``$B$'' superscript on the bare parton density in \eref{trackB} to distinguish it from the bare PDF of \eref{barepdf}.)
The next step in the factorization is to show that the partonic part can be written as the convolution of a finite hard coefficient and a separate, process-independent factor $Z^{\text{pdf},B}$ that contains all the collinear poles, 
\begin{equation}
\label{e.Zfactor}
\widehat{F}^{\text{partonic}} = \mathcal{C}^B \otimes Z^{\text{pdf},B} \, .
\end{equation}
We again use ``$B$'' superscripts on $\mathcal{C}^B$ and $Z^{\text{pdf},B}$ to distinguish these quantities from the analogous ones that appear in \eref{schematic} and \eref{renpdf}, respectively. 
Then substituting \eref{Zfactor} into \eref{trackB} and using the associativity of the convolution, we write
\begin{equation}
\label{e.trackBb}
F(x,Q^2) = \parz{\mathcal{C}^B \otimes Z^{\text{pdf},B}} \otimes f^{\text{bare},B} = \mathcal{C}^B \otimes \parz{Z^{\text{pdf},B} \otimes f^{\text{bare},B}} \, , 
\end{equation}
and define the last quantity in parentheses as a renormalized PDF,
\begin{equation}
\label{e.trackBc}
f^{\text{renorm},B}  =  Z^{\text{pdf},B} \otimes f^{\text{bare},B} \, .
\end{equation}
With this definition, \eref{trackBb} becomes a factorization formula in terms of a PDF 
$f^{\text{renorm},B}$, which shifts all the collinearly divergent but universal poles out of the partonic scattering matrix element $\widehat{F}^{\rm partonic}$ and combines them with $f^{\text{bare},B}$.
In this way, properties of the PDF that, in the operator-based approach, follow from UV renormalization of the Lagrangian (\sref{cons}) and composite operators (\sref{pdfs}) are instead determined from the behavior of collinear divergences in partonic structure function calculations. In this way, questions related to the UV renormalization of the fully non-perturbative PDF's operator definition get recast as issues related to infrared collinear divergences.

The main challenge in extending the track-B construction outlined above to fully general, non-perturbative PDFs is to identify precise definitions for each of the quantities with ``$B$''-superscripts \erefs{trackB}{trackBc} and to reconcile their properties with those of the fully non-perturbative operator-defined quantities in \sref{pdfs}.
A more detailed discussion of this, and a broader critique of track-B approaches, is given in Ref.~\cite[Sec.~III]{Collins:2021vke}. 
For our purposes, the relevant point is that, if \eref{trackB} is identified with the factorization formula \eref{schematic}, 
then the ``$B$''-superscripted quantities must match the corresponding quantities from \sref{pdfs},
\begin{align}
\label{e.ABmatch}
\begin{aligned}
\mathcal{C}^B & = \mathcal{C}(\mu;\epsilon) \, ,  \\
f^{\text{bare},B}   & = f_{0}(\xi;\epsilon) \, , \\
Z^{\text{pdf},B} & = Z^{\text{pdf}}(\mu;\epsilon) \, , \\
f^{\text{renorm},B} & =  f(\xi,\mu;\epsilon) \, ,
\end{aligned}
\end{align}
where the objects on the right sides are from \eref{schematic}, \eref{barepdf}, \eref{renpdf}, and \eref{barepdfunsub}. 
As explained in Sec. III A of Ref.~\cite{Collins:2021vke} (see also Ref.~\cite{Curci:1980uw}), this correspondence can be made to hold in ordinary minimal subtraction with $\mu = \mu_F$, that is, with all $A$'s in \erefs{ct1}{ct5} and \eref{ct6} equal. 
The same complications discussed in \sref{sumrulesa} reemerge, however, if separate $A$'s are used in the different types of pole subtractions while the usual sum rules and Ward identities continue to be imposed on the PDFs, as we illustrate in detail the next section.

A common strategy for deriving properties of $f^{\text{renorm},B}$  
is to show that $Z^{\text{pdf},B}$ preserves the given property under consideration while \emph{postulating} that it holds for $f^{\text{bare},B}$. 
Say that we wish to derive property $X$ for a track B ``renormalized'' PDF. 
We start by assuming that it holds for the track B bare PDF.
The renormalized PDF $f^{\text{renorm},B}$ then inherits property $X$ via its definition in \eref{trackBc}.
That strategy motivates an extremely generalized extension 
of a minimal subtraction scheme for PDFs, beyond even \eref{genMSreplace} and \eref{ctp}, wherein the only requirement is that the PDF counterterm 
$Z^{\text{pdf},B}$ takes the form, 
\begin{equation}
Z^{\text{pdf},B} = \delta(1-\xi) + \frac{1}{\epsilon}Z^{\text{pdf},B,(1)}(\epsilon;\mu;\mu_F;\dots)  + \frac{1}{\epsilon^2} Z^{\text{pdf},B,(2)}(\epsilon;\mu;\mu_F;\dots) + \hdots \, . \label{e.verygen}
\end{equation}
That is, the terms beyond the $\delta$-function are a completely general expansion in powers of $1/\epsilon$, where the $Z^{\text{pdf},B,(j)}$ coefficients are allowed to include any powers of the coupling $g$ and positive powers of the dimensional regulator $\epsilon$ as well as any number of multiple auxiliary scales, including $\mu$, $\mu_F$, and potentially others not explicitly written here. 
As long as $\mathcal{C}^B$ is defined so that \eref{Zfactor} holds and $f^{\text{renorm},B}$ is defined through \eref{trackBc}, then nothing about this setup prohibits such a broad generalization of the counterterm as in \eref{verygen}. 
There are no restrictions from  the properties of the PDF at the operator level.
Any prescription for moving contributions out of the hard part and regrouping them with $f^{\text{renorm},B}$ is allowed within such an approach, so long as the resulting hard part contains no divergent poles at $\epsilon = 0$.
Returning to the treatment of the number sum rule, if a specific prescription for the coefficients $Z^{\text{pdf},B,(i)}$ in \eref{verygen} is found satisfying
\begin{equation}
\label{e.zeroZ}
\int_0^1 \diff{\xi}{} Z^{\text{pdf},B,(i)}_{qj} = 0 \, 
\end{equation}
 for all $i \geq 1$, then we would have 
\begin{equation}
\int_0^1 \diff{\xi}{} Z_{qj}^{\text{pdf},B} = \delta_{qj} \, ,  
\end{equation}
analogous to \eref{Zint}. 
The sum rule for $f^{\text{renorm},B}$ is then recovered by applying the $\xi$-integral to \eref{trackBc} and by repeating exactly the steps of \eref{renpdf4}.
Hence, properties like the sum rules are essentially defined rather than derived when a factorization scheme is formulated to satisfy properties like \eref{zeroZ}.

By contrast with the flexibility afforded by the track-B approach above, the bare operator definition of the PDF in \eref{barepdf} imposes nontrivial constraints on the allowed form of the expansion in \eref{verygen}.
Additionally, although there is freedom in the choice of renormalization scheme and finite counterterms, RG invariance relates the different counterterms to one another and constrains their pole structure.
Therefore, fixing the $Z^{\text{pdf},B,(i)}$'s by \textit{fiat} in expansions like \eref{verygen} to enforce a desired set of properties will generally lead to results that are inconsistent with RG invariance. 
To demonstrate these points, we present some explicit examples in the next section using the Yukawa theory of \sref{cons}.

\section{Demonstrating sum rules and their breakdown when $\mu_F \neq \mu$}
\label{s.sumrulefail}

When the expected cancellations required to preserve sum rules (or other desired properties of PDFs) fail in perturbative QCD calculations, it is tempting to attribute the problem to difficulties with non-perturbative physics and to assume that it would be resolved in a more complete treatment of infrared or non-perturbative QCD.
However, the \sout{relevant} derivations are general consequences of UV renormalization in any finite-distance renormalizable theory, not just in QCD, and they do not depend on the presence of confinement or strong coupling in the infrared.
Therefore, violations of properties such as PDF sum rules cannot simply be attributed to non-perturbative behavior.
To isolate the effects of UV renormalization, it is useful to study theories where the infrared structure is sufficiently simple that it can be calculated explicitly, allowing us to verify which properties follow solely from the UV renormalization procedure.
We follow this strategy here by continuing to use the Yukawa theory specified in \eref{thelagrange}.
We confirm  by explicit calculation that the  standard quark number (\eref{renpdf4}) and momentum sum rules (\eref{momrule}) are indeed satisfied up to NLO in ordinary ${\rm MS}$ or $\msbar$ renormalization, but are violated in generalized minimal subtraction schemes with an independent factorization scale $\mu_F \neq \mu$. 

\subsection{Number and momentum sum rules}
\label{s.nummom}

\begin{figure}[t]
    \centering
    \includegraphics[width=15.0cm]{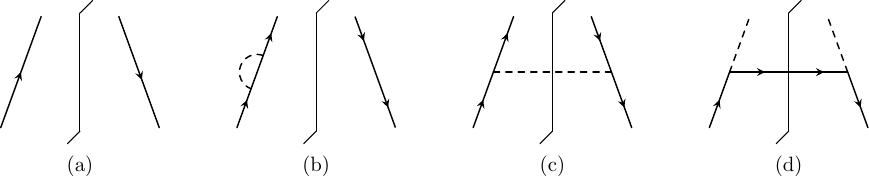}
    \caption{(a--c) $f_{q/q}^{\rm unsub}$ and (d) $f_{s/q}^{\rm unsub}$ up to $\order{a_g}$. Note that $(\rm b)$ does not show the hermitian conjugate diagram explicitly.}
    \label{f.LO-NLO_pdf}
\end{figure}

The lowest-order counterterms in Yukawa theory are obtained from the two- and three-point functions,
\begin{subequations}
\label{e.Zvalues}
\begin{align}
Z_s &= 1 - 2 a_g \frac{A_s^\epsilon}{\epsilon} + \order{a_g^2} 
\label{e.Zsvalue} \, , \\
Z_q &= 1 - \frac{a_g}{2} \frac{A_q^\epsilon}{\epsilon} + \order{a_g^2} 
\, , \label{e.Zqvalue} \\ 
\delta g &= \mu^\epsilon g \left[ a_g \frac{A^\epsilon}{\epsilon} + \order{a_g^2} \right] 
\, ,
\end{align}
\end{subequations}
where we have introduced the abbreviation, 
\begin{equation}
\label{e.agabbrev}
a_g = \frac{g^2}{(4 \pi)^2}
\end{equation}
and we allow for general subtraction coefficients as in \eref{genMSreplace},
\begin{equation}
A_q = \frac{\mu^2}{\mu_q^2} A\, , \quad A_s = \frac{\mu^2}{\mu_s^2} A \, .  \label{e.modA1}
\end{equation}
with $A = 4 \pi e^{-\gamma_E}$ being the default constant for the coupling renormalization. 
Likewise, directly calculating the PDFs through $\order{a_g}$ gives the counterterms
\begin{align}
Z_q Z_{qq}^\text{pdf} &= \delta_{qq} \delta(1-\xi) - a_g (1-\xi) \frac{A_{\text{pdf},qq}^\epsilon}{\epsilon} + \order{a_g^2} \, , \label{e.ZqqA} \\
Z_s Z_{sq}^\text{pdf} &= \delta_{sq} \delta(1-\xi) - a_g \xi \frac{A_{\text{pdf},sq}^\epsilon}{\epsilon} + \order{a_g^2} \, , \label{e.ZsqA}
\end{align}
where we define
\begin{equation}
A_{\text{pdf},qq} = \frac{\mu^2}{\mu_{\text{pdf},qq}^2} A  \, ,\quad A_{\text{pdf},sq} = \frac{\mu^2} {\mu_{\text{pdf},sq}^2} A \, . \label{e.modA2}
\end{equation}
Setting $\mu_q = \mu_s = \mu_{\text{pdf},qq} = \mu_{\text{pdf},sq} = \mu$ restores the ordinary $\msbar$ renormalzation scheme.

The lowest order contribution to the quark-in-quark PDF plus the leading virtual external leg corrections, represented by diagrams (a) and (b) of \fref{LO-NLO_pdf}, respectively, is obtained by applying the LSZ formalism, which gives
\begin{align}
& \; f^{\text{LO}}_{q/q}(\xi;\mu) + f^{\text{virt}}_{q/q}(\xi;\mu) = \parz{\sqrt{\text{residue}}}^2 \,  f^{\text{LO}}_{q/q}(\xi;\mu) \no
& =
\left[ 1 - \frac{a_g}{2} \left\{  \ln \parz{\frac{\mu^2}{m_q^2}} + 3 \frac{m_s^2}{m_q^2} + \left[\frac{3}{2} \frac{m_s^2}{m_q^2} \left(4 - \frac{m_s^2}{m_q^2}\right) - 4\right]\ln \parz{\frac{m_s^2}{m_q^2}}  - 7 \vphantom{\tanh^{-1} \left[ \sqrt{\frac{A}{A-4}} \right]} \right. \right. \no
& \hspace{1.75in} \left. \left. + 3 \left(2 - \frac{m_s^2}{m_q^2} \right) \frac{m_s}{m_q} \sqrt{4-\frac{m_s^2}{m_q^2}} \cos^{-1} \left(\frac{m_s}{2 m_q} \right) \right\}  - \frac{a_g}{2} \ln \parz{\frac{\mu_q^2}{\mu^2}} \right] \delta(1-\xi) \, . \label{e.qinqLOvirt}
\end{align}
We restrict to the case where $m_s < 2 m_q$.
The real emission quark-in-quark contribution from diagram (c) of \fref{LO-NLO_pdf} is
\begin{align}
& f^{\text{real}}_{q/q}(\xi;\mu) =
a_{g} (1-\xi) \left[ \ln \parz{\frac{\mu^2}{(1-\xi)^2 m_q^2 + \xi m_s^2}} + \frac{\xi \parz{4 m_q^2 - m_s^2}}{(1-\xi)^2 m_q^2 + \xi m_s^2} \right] + a_{g} (1-\xi) \ln \parz{\frac{\mu_{\text{pdf},qq}^2}{\mu^2}} \, . 
\label{e.thequarkpdf}
\end{align}
To verify the momentum sum rule, we will also need the scalar-in-quark PDF up to $\order{a_g}$, shown in diagram (d) \fref{LO-NLO_pdf}, 
\begin{align}
f_{s/q}(\xi;\mu) =
a_g \xi \left[ \ln \parz{\frac{\mu^2}{(1-\xi)m_s^2 + \xi^2 m_q^2}} + \frac{(1-\xi) \parz{4 m_q^2 - m_s^2}}{(1-\xi)m_s^2 + \xi^2 m_q^2} \right] + a_g \xi \ln \parz{\frac{\mu_{\text{pdf},sq}^2}{\mu^2}} \, . \label{e.scalarpdf}
\end{align}
At this order, the antiquark-in-quark PDF $f_{\bar{q}/q}$ vanishes and first receives contributions at $\order{a_g^2}$.
In \eref{qinqLOvirt}, \eref{thequarkpdf}, and \eref{scalarpdf}, the logarithms in the last terms are due to the modified $A$'s in \eref{modA1} and \eref{modA2}. 
Note that we leave the scale dependence of the renormalized coupling and masses implicit in the equations above for brevity. 

Performing the integration for the number sum rule in \eref{renpdf4} gives\footnote{For explicit analytic results, see \aref{integrals}.}
\begin{align}
\begin{aligned}
\int_0^1 \diff{\xi}{} \left[ f_{q/q}(\xi;\mu) - f_{\bar{q}/q}(\xi;\mu) \right] &= \int_0^1 \diff{\xi} \parz{f^{\text{LO}}_{q/q}(\xi;\mu) + f^{\text{virt}}_{q/q}(\xi;\mu) + f^{\text{real}}_{q/q}(\xi;\mu)} + \order{a_g^2} \\
& = 1 + \frac{a_g}{2} \ln \parz{\frac{\mu_{\text{pdf},qq}^2}{\mu_q^2}} + \order{a_g^2} \, ,
\end{aligned}
\label{e.numsumruletest}
\end{align}
and similarly, for the momentum sum rule \eref{momrule}, we have
\begin{align}
\sum_j \int_0^1 \diff{\xi} \xi f_{j/q}(\xi;\mu) &= \int_0^1 \diff{\xi} \xi \parz{f^{\text{LO}}_{q/q}(\xi;\mu) + f^{\text{virt}}_{q/q}(\xi;\mu) + f^{\text{real}}_{q/q}(\xi;\mu) + f_{s/q}(\xi;\mu)} + \order{a_g^2} \, \no
&= 1 + \frac{a_g}{6} \ln \parz{\frac{\mu_{\text{pdf},qq}^2}{\mu_{\text{pdf},sq}^2}} + \frac{a_g}{2} \ln \parz{\frac{\mu_{\text{pdf},sq}^2}{\mu_q^2}} + \order{a_g^2}  \, .
\label{e.momsumruletest}
\end{align}
To preserve the number sum rule, \eref{numsumruletest} shows that the quark PDF scale needs to be set equal to the quark wavefunction renormalization scale, $\mu_q = \mu_{\text{pdf},qq}$.
Separately, preserving the momentum sum rule requires that the quark renormalization, scalar PDF, and quark PDF scales are all equal.
Thus, simultaneously preserving both sum rules requires that all scales are set equal to one another 
\begin{equation}
\mu_q = \mu_{\text{pdf},qq} = \mu_{\text{pdf},sq} \, . 
\end{equation}

Notice that the cancellation of UV divergences  necessary to maintain the sum rules takes place between rather different types of diagrams.
In \eref{qinqLOvirt}, the UV divergences arise from quark self-energy graphs, and the renormalization is just the ordinary wavefunction renormalization from the Lagrangian, while in \erefs{thequarkpdf}{scalarpdf} the UV divergences are from the extra real emission graphs in the unsubtracted PDF, where the renormalization involves the extra $Z^{\text{pdf},qq}$ and $Z^{\text{pdf},sq}$ counterterms. 
That the sum rules depend on cancellations between the $Z_q$ and the $Z^{\text{pdf},qq(sq)}$ counterterms highlights the non-trivial relationship between ordinary renormalization scales and factorization scales.

\begin{figure}[t]
    \centering
    \includegraphics[width=13cm]{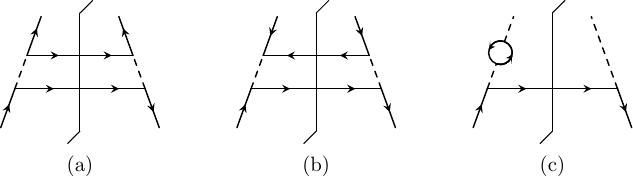}
    \caption{Example graphs that require $\mu_q \!=\! \mu_s$ for the momentum sum rule to hold at $\order{a_g^2}$. Diagrams contributing to (a) $f_{q/q}^{\rm unsub}$, (b) $f_{\bar{q}/q}^{\rm unsub}$, and (c) $f_{s/q}^{\rm unsub}$ at $\order{a_g^2}$. Note that (c) does not show the hermitian conjugate diagram explicitly.}
    \label{f.NNLO_pdf}
\end{figure}

Considering the next order in $a_g$, it becomes clear that the preservation of the sum rules also requires $\mu_s = \mu_q$. 
For the momentum sum rule, it is the sum of graphs (a) and (b) in \fref{NNLO_pdf} that cancel when $\mu_s = \mu_{\text{pdf},qq} = \mu_{\text{pdf},sq}$ in the same way as the $\order{a_g}$ example with the graphs in \fref{LO-NLO_pdf}.
To summarize, all the wavefunction and PDF renormalization scales need to be set equal for both the number and momentum sum rules to be preserved simultaneously, 
\begin{equation}
\mu_s = \mu_q = \mu_{\text{pdf},qq} = \mu_{\text{pdf},sq} \, . 
\end{equation}

\subsection{Gauge invariance}
\label{s.gaugeinv}

The discussion in \sref{nummom} above deals only with the connection between the PDF and wavefunction renormalization scales in the derivation of PDF sum rules. 
However, similar arguments apply to Ward identities in gauge theories. 
As explained, for example, in Ref.~\cite[pgs. 384-385]{Sterman:1993hfp}, preserving gauge invariance order-by-order in perturbation theory requires the $A$ constants in the quark wavefunction and coupling renormalization factors (\eref{cts}) to be equal, and therefore the corresponding wave function renormalization scale $\mu_q$ and gauge coupling scale $\mu_e$, in both the abelian and non-abelian cases.

To transplant the Ward identity argument into the Yukawa theory example used above, we append a Maxwell field term to the Lagrangian in \eref{thelagrange},
\begin{equation}
\begin{aligned}
 \mathcal{L}_\text{Maxwell} &= -\frac{1}{4} F_{0,{\mu \nu}}F_0^{\mu \nu} - e_0 \overline{\psi}_{q,0} \gamma_\mu \psi_{q,0} A_{0}^\mu - \frac{1}{2 \zeta_0} (\partial_\mu A_0^\mu)^2 \\
 &= -\frac{Z_\gamma}{4} F_{{\mu \nu}}F^{\mu \nu} - e_0 Z_q Z_\gamma^{1/2}\overline{\psi}_{q} \gamma_\mu \psi_{q} A^\mu - \frac{Z_\gamma}{2 \zeta_0} (\partial_\mu A^\mu)^2\, \\
 &= -\frac{1}{4} F_{{\mu \nu}}F^{\mu \nu} - \mu_e^\epsilon e \overline{\psi}_{q} \gamma_\mu \psi_{q} A^\mu - \frac{1}{2 \zeta} (\partial_\mu A^\mu)^2 -\frac{\delta Z_\gamma}{4} F_{{\mu \nu}}F^{\mu \nu} -  \delta e \, \overline{\psi}_{q} \gamma_\mu \psi_{q} A^\mu \, , 
\end{aligned}
\end{equation}
where $\zeta$ is the usual Lorenz gauge fixing parameter, and on the second and third lines we have made the usual renormalization substitutions, 
\begin{align}
 A_0^\mu = Z_\gamma^{1/2} A^\mu \, , \qquad  Z_\gamma &= 1 + \delta Z_\gamma \, ,  \qquad   Z_q Z_\gamma^{1/2} e_0 = \mu_e^\epsilon e + \delta e \, , \qquad \zeta_0 = \zeta Z_\gamma \, .
\end{align}
In generalized minimal subtraction schemes
\begin{align}
 \delta e(e(\mu_e),g(\mu),\epsilon) &=  \mu_e^\epsilon e(\mu_e) \sum_{n=1}^\infty \parz{\frac{A_e^\epsilon}{\epsilon}}^n C^e_n(e(\mu_e),g(\mu)) \, ,  \label{e.ctem1} \\
\delta Z_\gamma(\mu_e,\mu,\epsilon) &= \sum_{n=1}^\infty \parz{\frac{A_\gamma^\epsilon}{\epsilon}}^n C^{Z_\gamma}_n(e(\mu_e),g(\mu)) \label{e.ctem5} \, .
\end{align}
\begin{figure}[t]
    \centering
    \includegraphics[width=8.5cm]{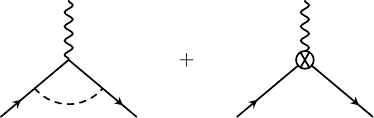}
    \caption{The $\order{a_g}$ gauge field vertex correction with counterterm.}
    \label{f.vertex}
\end{figure}
Calculating the $\order{a_g}$ gauge field counterterm from the lowest order amputated three-point graph (\fref{vertex}) gives
\begin{align} 
\frac{\delta e}{\mu_e^\epsilon e} \equiv Z_1 -1 = -\frac{a_g(\mu)}{2} \frac{A_e^\epsilon}{\epsilon} + \cdots \, , \label{e.emvertct}
\end{align}
where $Z_1$ denotes the electromagnetic interaction vertex renormalization to match the notation from the discussion in  Ref.~\cite[pg. 335]{Sterman:1993hfp}.
Gauge invariance imposes the counterterm Ward identity $Z_1 = Z_q$, which is required to preserve the Feynman identity in~\cite[Fig.~11.5]{Sterman:1993hfp}, and comparing \eref{Zqvalue} with \eref{emvertct} shows that this Ward identity is satisfied through $\order{a_g}$ in our Yukawa theory example only if
\begin{equation}
A_q = A_e \, .
\end{equation}

The discussion above emphasizes that, similar to the case with the sum rules from \sref{nummom}, gauge symmetry Ward identities are preserved because of cancellations between different types of counterterms. 
In the case of the Feynman identity, it is a cancellation between the wavefunction renormalization and the gauge coupling renormalization. 
Preserving Ward identities in gauge theories is essential because they constrain the gauge-invariant operator structure of factorization, including the Wilson lines appearing in non-perturbative definitions of PDFs.
Altering these identities through additional independent factorization scales can therefore modify the operator-level formulation underlying factorization and its connection to PDFs.

\subsection{Effect of renormalizing the coupling and wavefunction separately}

The only coupling present in QCD is the gauge coupling, so simultaneously preserving the PDF sum rules and Ward identities forces us to use a single universal subtraction constant ``$A$'', or equivalently, a single renormalization scale $\mu$.
Nevertheless, it is instructive to consider the effect of RG invariance in a non-gauge theory such as the Yukawa theory in \eref{thelagrange} when there are separate renormalization and factorization scales for the coupling counterterm (\eref{ct5}) and other counterterms (\erefs{ct1}{ct4}), respectively.
To that end, let us now return to the pure scalar Yukawa theory but take
\begin{equation}
\mu_s = \mu_q = \mu_{\text{pdf},qq} = \mu_{\text{pdf},sq} \equiv \mu_F \ne \mu \, .
\end{equation}
That is, all scales apart from the scale from the coupling renormalization are equal to each other. 
Since $\mu_F$ is an unphysical renormalization scheme parameter, RG invariance implies that the renormalized coupling cannot be treated as independent of the choice of $\mu_F$.
To see this, note that the expansion of $g_0$ in terms of the renormalized coupling $g$, with $g$ held fixed, depends on $\mu_F$ through the quark and scalar wavefunction renormalization from \erefs{renorms}{Zs},
\begin{equation}
Z_q Z_s^{1/2} g_0 = \mu^\epsilon g + \delta g \, .
\end{equation}
Because of the additional dependence on $\mu_F$ that enters through the field renormalization factors, the full running of the coupling follows from the pair of RG equations
\begin{subequations}
\label{e.rgg}
\begin{align}
&  \frac{\diff{g_0}{}}{\diff{\ln \mu}{}} = 0 \, , \label{e.rgg01} \\
&  \frac{\diff{g_0}{}}{\diff{\ln \mu_F}{}} = 0 \, , \label{e.rgg02}
\end{align}
\end{subequations}
which can be used to obtain the dependence of the renormalized coupling $g(\mu,\mu_F)$ on both $\mu$ and $\mu_F$ through the $\beta$-functions
\begin{subequations}
\label{e.betafuncs}
\begin{align}
\widetilde{\beta}  &\equiv \frac{\partial g(\mu,\mu_F)}{\partial \ln \mu} = \frac{g^3}{8 \pi^2} + \order{g^5} \, , \\ 
\widetilde{\beta}_F  &\equiv  \frac{\partial g(\mu,\mu_F)}{\partial \ln \mu_F} = \frac{3 g^3}{16 \pi^2}  + \order{g^5} \, ,
\end{align}
\end{subequations}
where the last set of equalities comes from solving \eref{rgg} with the lowest order graphs.
We can recover a single-scale running coupling if we treat $\mu_F$ as a function of $\mu$ so that the full $\beta$-function is
\begin{equation}
\label{e.fullbeta}
\beta(\mu,\mu_F(\mu)) \equiv \frac{\diff g}{\diff \ln{\mu}} = \widetilde{\beta} + \widetilde{\beta}_F \frac{\diff{\ln \mu_F(\mu)}{}} {\diff{\ln \mu}{}}  \, .
\end{equation}
The usual $\msbar$ coupling and $\beta$-function are recovered when $\mu_F(\mu) = \mu$. 

We can rewrite \eref{betafuncs} in favor of $a_g$, which gives
\begin{subequations}
\begin{align}
\frac{\partial a_g(\mu,\mu_F) }{\partial \ln \mu} &= 4 a_g^2 + \order{a_g^4} \, , \\
\frac{\partial a_g(\mu,\mu_F) }{\partial \ln \mu_F} &= 6 a_g^2 + \order{a_g^4} \, . \label{e.mFevol}
\end{align}
\end{subequations}
Using the initial condition $a_g^{\msbar}(\mu) \equiv a_g(\mu,\mu_F = \mu)$ to solve for the lowest order  $\mu_F$-running from \eref{mFevol} gives
\begin{equation}
a_g(\mu,\mu_F) 
= \frac{a^{\msbar}_g(\mu)}{1-6 a^{\msbar}_g(\mu) \ln (\mu_F / \mu)} = a^{\msbar}_g(\mu) + 6 \big[ a^{\msbar}_g(\mu) \big]^2 \ln \parz{\frac{\mu_F}{\mu}} + \order{\big[a_g^{\msbar}\big]^3} \, . \label{e.couplingcorrect}
\end{equation}
We see that it is not possible to introduce an additional, independent factorization scale $\mu_F \ne \mu$ while simultaneously enforcing the PDF sum rules and keeping the numerical value of the coupling fixed at its $\msbar$ value.
Either the sum rules need to be modified with correction terms or the coupling must to be shifted, including in the PDFs of \sref{nummom} and the renormalization factors of the Lagrangian and PDFs in \eref{Zvalues} and \erefs{ZqqA}{ZsqA}, respectively.
Furthermore, because of this, the RG evolution of the PDFs becomes governed by a set of coupled equations, whether the corresponding sum rules are enforced or not, for the dependence on the renormalization scale $\mu$ and each additional factorization scale.
We suspect that this is another way to make the similar points that appear in Sec.~IV of Ref.~\cite{Harland-Lang:2018bxd}.
These evolution equations are obtained by solving the RG invariance relations, analogous to \eref{rgg} for the coupling,
\begin{subequations}
\label{e.extdglap}
\begin{align}
\frac{\diff{}}{\diff{\ln \mu}{}} f_{j/q,0}(\xi;\epsilon) &= 0 \, , \label{e.pdfrgp1} \\
\frac{\diff{}}{\diff{\ln \mu_F}{}} f_{j/q,0}(\xi;\epsilon) &= 0 \, . \label{e.muFdglap}
\end{align}
\end{subequations}
Therefore, to avoid any ambiguities in the treatment of evolution, all auxiliary scales should be listed as explicit arguments in both the PDFs $f_{j/q}(\xi;\mu;\mu_F)$ in \sref{nummom} and in the coupling. 

These observations seem to contradict earlier results that do use multiple scales, and it is important to understand how both can be valid simultaneously. 
For example, Appendix A of Ref.~\cite{Chen:2016zka} argues, starting from the usual single-scale $\msbar$ coupling, that multi-scale PDF evolution is equivalent to using a single factorization scale for evolution, with the only difference being a shift in the scale of the coupling and, possibly, the value of the PDF at the initial scale.
What our results show is that such a method for removing multiscale evolution is only consistent with RG invariance if the symmetries that give rise to sum rules are discarded at the level of the fully non-perturbative definitions for the renormalized PDFs.

\section{Implications for phenomenological applications}
\label{s.pheno}

In the precision era of high-energy and nuclear physics, there is significant impetus to improve the precision of theoretical inputs to experimental efforts, both from the non-perturbative functions such as PDFs and the accuracy of perturbative expansions.
These are particularly important for the determination of fundamental Standard Model (SM) physics parameters, for which theoretical uncertainties remain a dominant limitation on precision~\cite{dEnterria:2022hzv,Cerci:2023uhu,Gao:2017yyd}.
Furthermore, simultaneous extractions of these parameters, especially the strong coupling $\alpha_s$ at the mass of the $Z$-boson, alongside PDFs are becoming standard practice~\cite{Ablat:2025gbp,Ball:2018iqk,Ball:2025xgq,Cridge:2021qfd,Alekhin:2017kpj}.
In light of \sref{sumrulefail}, care must be taken when extracting SM parameters phenomenologically because the renormalization procedure can influence their determination. 
For example, we observed that imposing the usual PDF sum rules in generalized minimal subtraction, while introducing a factorization scale separate from the renormalization scale, means that the renormalized coupling and mass depend on two scales, and initial-scale values of such parameters differ from the corresponding single-scale values.
If data are used across a large range of energy scales, as is the case in standard global PDF analyses, the modified scale dependence of the coupling away from the $Z$-pole may introduce biases that must be compensated by adjusting its value at the initial scale.
Generally, the most straightforward and internally consistent procedure is to employ a single renormalization group parameter $\mu$ throughout such analyses.

Even when a single scale $\mu$ is used to define central values in phenomenological fits to data, it is common to assign theoretical uncertainties from missing higher-order uncertainties (MHOUs) by independently varying factorization and renormalization scales.
The most common prescription is to evaluate $\Gamma(\mu\!=\!r Q,\mu_F\!=\!r_F Q)$, where $\Gamma$ is a generic observable and $Q$ denotes a hard scale of the process under study, with $r,\, r_F \in \{ 1/2,1,2 \}$, and to assign the MHOU as the maximum deviation from the central prediction~\cite{ParticleDataGroup:2008zun,Harland-Lang:2018bxd,LHCHiggsCrossSectionWorkingGroup:2016ypw}.
More sophisticated methods for defining MHOUs also exist and consider correlations between different kinematics and processes under scale variation~\cite{NNPDF:2019ubu,NNPDF:2024dpb,Bagnaschi:2014wea,Cacciari:2011ze,Bonvini:2020xeo,Tackmann:2024kci}.
What our results emphasize is that such variations need to also account for the separate $\mu$ and $\mu_F$ variations in the coupling and masses, as well as the impact of possible gauge invariance violations. 

Accurate theoretical uncertainty estimates are essential for meaningful comparisons between theory and experiment and for reliably assessing the precision of non-perturbative 
PDFs or fundamental SM parameter extractions.
In exact quantities, there is no sensitivity to auxiliary scales. 
All scale sensitivity in the hard part must be exactly canceled by an opposite sensitivity in the PDF, so, at fixed order, any residual auxiliary scale sensitivity is due to the neglect of higher orders in the hard coefficient.
Therefore, one way to estimate the uncertainties due to neglected higher orders is to vary auxiliary scales in the untruncated hard part and observe the size of the impact. 

As a concrete example, consider the inclusive $F_2$ structure function on a proton target~\cite{CTEQ:1993hwr}, calculated to $\order{\alpha_s^n}$,
\begin{align}
    F_{2}(x,Q) = \sum_{i} \int_{x}^{1} \diff{\xi} \, \mathcal{C}^{(n)}_{2,q/i}\Big( \frac{x}{\xi},\frac{\mu}{Q},\frac{\mu_F}{\mu}, \zeta \Big) \underline{f}^{(n)}_{i/P}(\xi;\mu;\mu_F;\zeta) + \order{\alpha_s^{n+1}} 
    + \order{ \frac{m^2}{\mu^2} }
\label{e.F2-factorized}
,\end{align}
where we denote the first term by $F_{2}^{(n)}(x,Q;\mu,\mu_F,\zeta)$ below to emphasize that it is the factorized approximation to the structure function.
The superscript ``$(n)$'' indicates consistent truncation of the perturbative expansion at $\order{\alpha_s^{n}}$.
Additionally, the underline on the PDF distinguishes the phenomenological PDFs obtained from data from the idealized ``ground-truth'' non-perturbative PDFs such as those defined in \sref{pdfs}.
As discussed in \sref{gaugeinv}, general multiscale schemes can violate gauge invariance of the PDF, so we also include a parameter $\zeta$ to indicate a potential continuous gauge-fixing parameter.   
Through $\order{\alpha_s^n}$, any variations in $\mu$ or $\mu_F$ cancel exactly between $\mathcal{C}^{(n)}_{2,q/i}$ and $\underline{f}^{(n)}_{i/q}$, and likewise, any dependence on the gauge fixing parameter $\zeta$ must cancel through order $\alpha_s^n$. 
However, residual scale dependence remains at any finite $n$ because of the neglected $\order{\alpha_s^{n+1}}$ corrections in \eref{F2-factorized}. 
To keep track of all the different contributions to auxiliary scale dependence, it is important that all factors be expressed as functions of both $\mu$ and $\mu_F$, emphasized recently in Ref.~\cite{Hampson:2025pvi}.

As mentioned above, one way to estimate uncertainties from neglected higher orders is to vary the auxiliary scales and observe the size of the impact on the approximated structure function $F_2^{(n)}$. 
In RG schemes with multiple auxiliary scales, doing this is complicated by the fact that the effects of changing $\mu$ and $\mu_F$ are intertwined in rather complex ways.
The example of the multiscale running coupling in \eref{couplingcorrect} makes this point: the impact of varying the ratio $\mu_F/\mu$ itself depends on the absolute size of $\mu$. 
Therefore, for testing multi-scale sensitivity in practice it is most straightforward to take $\mu_F$ to be a general function of $\mu$.
Then, we may apply 
\begin{equation}
\frac{\diff{}}{\diff{\ln}{\mu}} = \frac{\partial}{\partial \ln \mu} + \frac{\diff{\ln \mu_F(\mu)}}{\diff{\ln \mu}{} } \frac{\partial}{\partial \ln \mu_F}
\end{equation}
to \eref{F2-factorized}, and the only nonvanishing contributions will be from the $\order{\alpha_s^{n+1}}$ errors. 
That is, we calculate
\begin{align}
\frac{\diff{F_{2}^{(n)}}{}}{\diff{\ln \mu}{}} &= \sum_{i} \left[ \frac{\diff{\mathcal{C}^{(n)}_{2,q/i}}{}}{\diff{\ln \mu}{}} \otimes \underline{f}^{(n)}_{i/q} + \mathcal{C}^{(n)}_{2,q/i} \otimes \frac{\diff{\underline{f}^{(n)}_{i/q}}{}}{\diff{\ln \mu}} \right] \, . \label{e.F2der}
\end{align}
One should expect a useful scheme to be one corresponding to a path along the $(\mu,\mu_F)$-plane in which \eref{F2der} is always very small. 
Auxiliary scale sensitivity has long been used as a proxy for sensitivity to higher orders~\cite{Stevenson:1981vj,Stevenson:2019ytx,Politzer:1981vc,Stevenson:1986cu,Maxwell:2000mm,Wu:2019mky,Wu:2018cmb,Ma:2015dxa,Brodsky:1982gc,Deur:2017cvd}, and the above demonstrates how to generalize this to multiple scales. 
A similar formula is needed to test sensitivity to the gauge fixing parameter $\zeta$. 

Equation~\eqref{e.F2der} also provides a convenient way to determine the coefficient functions $\mathcal{C}^{(n)}_{2,q/i}$ in the multiscale factorization scheme. 
The coefficient functions on the line $\mu_F = \mu$ coincide with those of the ordinary single scale  minimal subtraction scheme, and for arbitrary $\mu$ and $\mu_F$, the coefficient functions may be obtained by integrating the evolution equation,
\begin{equation}
\label{e.coeffevol}
\frac{\diff{\mathcal{C}^{(n)}_{2,q/i}}{}}{\diff{\ln \mu}{}} = - \sum_j \mathcal{C}^{(n)}_{2,q/j} \otimes P^{(n)}_{j/i} \, ,
\end{equation}
implied by \eref{F2der}, along a chosen trajectory $\mu_F = \mu_F(\mu)$, through $\order{\alpha_s^n}$.
Note that $P^{(n)}(\xi;\mu_F/\mu;\zeta)$ is a multi-scale $\order{\alpha_s^{n+1}}$ perturbative evolution kernel, analogous to the multi-scale $\beta$-function in \eref{fullbeta}. 
Once \eref{coeffevol} has been solved and the full $\mu$ and $\mu_F$ dependence of the coefficient function has been determined, the effect of varying the renormalization scale $\mu$ and factorization scale along the path $\mu_F(\mu)$ in the hard part can be examined directly.

\section{Summary and conclusions}
\label{s.conclusions}

Summarizing the last few sections, we have verified the following for generalized minimal subtraction schemes:
\begin{enumerate}
\item[1)] In general, for quantum field theories that require renormalization, the  renormalization scale for the PDFs must (at least) equal the scales of the regular wavefunction renormalization to preserve the sum rules, including the number and momentum sum rules. 
\item[2)] In theories with a gauge symmetry, preserving the Ward identities implies that the scale of the wavefunction renormalization must also equal the scale used to renormalize the gauge coupling.
\item[3)] In multi-scale renormalization schemes, RG invariance imposes a non-trivial wavefunction renormalization scale dependence on the coupling since the wavefunction and coupling renormalization scales cannot be adjusted independently.

\item[4)] Altogether, therefore, the advantages of keeping constraints from RG invariance, PDF sum rules, Ward identities, and a single-scale evolution for the coupling suggest that it is preferable to keep all auxiliary scales equal.  
\end{enumerate}
It is always possible to introduce any number of arbitrarily many factorization scales $\mu_F$ into the phenomenological analyses of PDFs since they are by definition auxiliary scales that do not, in principle, affect physical observables particularly if properties of the PDF such as sum rules are not strictly imposed.
The corresponding scheme changes require that corrections be made to constraints like the sum rules that are related to underlying symmetries. 
For example, Ref.~\cite[see Eq.~(46)]{Delorme:2025teo} points out that switching factorization schemes can modify the PDF sum rules like \eref{renpdf4} such that
\begin{align}
& \int_0^1 \diff{\xi}{} \left[ f_{q/q}(\xi;\mu;\mu_F;\epsilon) - f_{\bar{q}/q}(\xi;\mu;\mu_F;\epsilon) \right] = 1 + \mathcal{H}  \otimes f \, 
 \label{e.renpdf4conc}
\end{align}
where $\mathcal{H}$ is a perturbatively calculable, scheme-dependent coefficient function. 
In this paper, we have shown that the need for such corrections extends to any generalization of dimensional analysis and pole subtraction scheme that introduces a separate factorization scale independent from the renormalization scale. 

Because of the points enumerated above, varying a $\mu_F$ relative to a fixed $\mu$ without simultaneously taking into account of the impact on other constraints could   misrepresent the sizes of uncertainties.
Even if sum rule corrections like \eref{renpdf4conc} are accounted for, corrections to the strong coupling and any constraints based on Ward identities also need to be estimated and taken into account.

To keep the discussion simple, we have focused on totally inclusive DIS because it involves only a single non-perturbative correlation function, the PDF. 
Exactly the same considerations apply, and are amplified, in more complex processes where multiple types of correlation functions are present at the same time.  For example, semi-inclusive DIS factorization formulas~\cite{Graudenz:1994dq,Goyal:2023zdi} are often expressed in the form, 
\begin{equation}
\label{e.schematic0}
F(x,z,Q) = \mathcal{C}(x/\xi,z/\zeta,Q/\mu;g(\mu),\mu_F,\mu_D) \otimes f(\xi;\mu_F) \otimes D(\zeta;\mu_D)
,\end{equation}
where here $D(\zeta;\mu_D)$ is a fragmentation function, and $\mu_D$ is a separate factorization scale associated with fragmentation. 
See, for example, Ref.~\cite[Eq.~(3)]{Bonino:2024qbh}. 
In most applications, all these auxiliary scales are ultimately equated to each other numerically in the end. 
What we have shown in this paper is that doing this is a requirement of factorization, renormalization group invariance, and the preservation of sum rules and underlying symmetries of the theory. 

Beyond phenomenological applications, the observations from this paper are relevant for methods of describing PDFs and other non-perturbative correlation functions that use non-perturbative techniques like lattice QCD~\cite{Constantinou:2020hdm,Radyushkin:2019mye,Ji:2013dva,Ji:2014gla,Ma:2014jla}. 
There, the connection to universal and interpretable matrix element definitions is central. 
In the case of lattice calculations, sum rule constraints appear in the power expansions that relate lattice-calculable objects like quasi- and pseudo-PDFs to lightcone PDFs. 
Further study on the impact of multiple auxiliary scales in this context is warranted. 

\appendix

\section{Integrals for checking sum rules}
\label{a.integrals}

For checking the sum rules, we need the following 
$\xi$-weighted integrals of the expressions from \sref{nummom}, 
\begin{align}
\int_0^1 \diff{\xi} & \parz{f^{\text{LO}}_{q/q}(\xi;\mu) + f^{\text{virt}}_{q/q}(\xi;\mu)} = \int_0^1 \diff{\xi} \xi \parz{f^{\text{LO}}_{q/q}(\xi;\mu) + f^{\text{virt}}_{q/q}(\xi;\mu)} \no 
& 1 - \frac{a_g}{2} \left\{  \ln \parz{\frac{\mu^2}{m_q^2}} + 3 \frac{m_s^2}{m_q^2} + \left[\frac{3}{2} \frac{m_s^2}{m_q^2} \left(4 - \frac{m_s^2}{m_q^2}\right) - 4\right]\ln \parz{\frac{m_s^2}{m_q^2}}  - 7 \vphantom{\tanh^{-1} \left[ \sqrt{\frac{A}{A-4}} \right]} \right.  \no
& \qquad \left. + 3 \left(2 - \frac{m_s^2}{m_q^2} \right) \frac{m_s}{m_q} \sqrt{4-\frac{m_s^2}{m_q^2}} \cos^{-1} \left(\frac{m_s}{2 m_q} \right) \right\} -\frac{a_g}{2} \ln \parz{\frac{\mu_q^2}{\mu^2}} \,  \label{e.virtualnumber} \, , \\
\int_0^1 \diff{\xi}  f^{\text{real}}_{q/q}(\xi;\mu) =& \frac{a_g}{2} \left\{  \ln \parz{\frac{\mu^2}{m_q^2}} + 3 \frac{m_s^2}{m_q^2} + \left[\frac{3}{2} \frac{m_s^2}{m_q^2} \left(4 - \frac{m_s^2}{m_q^2}\right) - 4\right]\ln \parz{\frac{m_s^2}{m_q^2}}  - 7 \vphantom{\tanh^{-1} \left[ \sqrt{\frac{A}{A-4}} \right]} \right.  \no
& \hspace{.4in} \left. + 3 \left(2 - \frac{m_s^2}{m_q^2} \right) \frac{m_s}{m_q} \sqrt{4-\frac{m_s^2}{m_q^2}} \cos^{-1} \left(\frac{m_s}{2 m_q} \right) \right\} + \frac{a_{g}(\mu)}{2} \ln \parz{\frac{\mu_{\text{pdf},qq}^2}{\mu^2}} \, , \label{e.realqqcalc} \\
\int_0^1 \diff{\xi} \xi f^{\text{real}}_{q/q}(\xi;\mu) =& 
\frac{a_g}{36} \left\{ 6 \ln \parz{\frac{\mu^2}{m_q^2}} + 6  \frac{m_s}{m_q} \sqrt{\parz{4-\frac{m_s^2}{m_q^2}}} \parz{8 \frac{m_s^4}{m_q^4} - 35 \frac{m_s^2}{m_q^2} + 30} \cos^{-1}\parz{\frac{m_s}{2 m_q}}  \right. \no
& \left. \qquad + 3 \parz{8 \frac{m_s^6}{m_q^6} - 51 \frac{m_s^4}{m_q^4} + 84 \frac{m_s^2}{m_q^2} - 24} \ln \parz{\frac{m_s^2}{m_q^2}} - 48 \frac{m_s^4}{m_q^4} + 234 \frac{m_s^2}{m_q^2} - 206 \vphantom{\cos^{-1}\parz{\frac{\sqrt{a}}{2}}} \right\} \no
& \qquad + \frac{a_g}{6} \ln \parz{\frac{\mu_{\text{pdf},qq}^2}{\mu^2}} \, , \label{e.quarkweighted} \\
\int_0^1 \diff{\xi} \xi f_{s/q}(\xi;\mu) =& \frac{a_g}{36} \left\{-12 \frac{m_s}{m_q} \sqrt{\parz{4-\frac{m_s^2}{m_q^2}}}\parz{4 \frac{m_s^4}{m_q^4} - 13 \frac{m_s^2}{m_q^2} +6}\cos^{-1}\parz{\frac{m_s}{2 m_q}} \right. \no
& \qquad  \left. -6 \frac{m_s^2}{m_q^2} \parz{4 \frac{m_s^4}{m_q^4} -21 \frac{m_s^2}{m_q^2} + 24} \ln \parz{\frac{m_s^2}{m_q^2}}   + 48 \frac{m_s^4}{m_q^4} -180 \frac{m_s^2}{m_q^2} + 80 \vphantom{\cos^{-1}\parz{\frac{\sqrt{a}}{2}}} \right\} \no
& \qquad - \frac{a_g}{6} \ln \parz{\frac{\mu_{\text{pdf},sq}^2}{\mu^2}} + \frac{a_g}{2} \ln \parz{\frac{\mu_{\text{pdf},sq}^2}{\mu^2}} \, . \label{e.scalarweighted}
\end{align}
In all cases, we drop $\order{a_g^2}$ errors. Generally, all couplings, masses and PDFs should be viewed as functions of both $\mu$ and $\mu_F$, but we leave this implicit to keep the formulas compact.

\vskip 0.3in
\begin{acknowledgments}
We thank John Collins, Joseph Karpie, Kazuki Makino, and James Whitehead for useful comments on the text. 
T. Rogers was supported by the U.S. Department of Energy, Office of Science, Office of Nuclear Physics, under Award Number DE-SC0024715.
manuscript.
The work of R.~M.~W. was partially supported by the Jefferson Science Associates (JSA) Graduate Fellowship.
This material is based upon work supported by the U.S. Department of Energy, Office of Science, Office of Workforce Development for Teachers and Scientists, Office of Science Graduate Student Research (SCGSR) program. The SCGSR program is administered by the Oak Ridge Institute for Science and Education (ORISE) for the DOE. 
ORISE is managed by ORAU under contract number
DESC0014664. 
All opinions expressed in this paper are the author’s and do not necessarily reflect the policies and views of DOE, ORAU, or ORISE. No AI was used in writing this article. 
\end{acknowledgments}

\bibliography{bibliography}

\end{document}